\documentclass[showpacs,final,a4paper,pra,10pt]{revtex4}

\begin{document}
\title{Dense coding quantum key distribution based on continuous variable
entanglement}
\author{Xiaolong Su, Jietai Jing, Qing Pan$^{*}$ and Changde Xie}
\affiliation{The State Key Laboratory of Quantum Optics and Quantum Optics\\
Devices, Institute of Opto-Electronics, Shanxi University, Taiyuan, 030006,\\
P. R. China}

\begin{abstract}
We proposed a scheme of continuous variable quantum key
distribution, in which the bright Einstein-Podolsky-Rosen entangled
optical beams are utilized. The source of the entangled beams is
placed inside the receiving station, only a half of the entangled
beams is transmitted with round trip and the other half is retained
by the receiver. The amplitude and phase signals modulated on the
signal beam by the sender are simultaneously extracted by the
authorized receiver with the scheme of the dense coding correlation
measurement for continuous quantum variables, thus the channel
capacity is significantly improved. Two kinds of possible
eavesdropping are discussed. The mutual information and the secret
key rates are calculated and compared with that of unidirectional
transmission schemes.

\end{abstract}

 \maketitle
\section{Introduction}

The quantum key distribution (QKD) based on the fundamental properties of
quantum mechanics provides a way for two distant parties, usually named
Alice (sender) and Bob (receiver), to share keys for encryption that can be
absolutely secret in principle. Generally, we say, a communication channel
is secure in the sense that any eavesdropper can be detected by the
authorized communication partners. Initially, the discussions and
experimentally demonstrations on QKD were concentrated within the setting of
discrete variables (dv). There have been two basic schemes for qubit-based
dv QKD, which are the sending of states from non-orthogonal bases, such as
the original ''BB84 protocol'' proposed in 1984 by Bennett and Brassard \cite
{one}, and those based on sharing entanglement between the sender and
receiver, such as Ekert's scheme (E91) \cite{two}. So far, a plenty of dv
QKD schemes have been presented and its potential application in the long
distance secure communication have been experimentally demonstrated \cite
{three,four,five,six,seven}. However, the data transmission rates and the
detection efficiency of single photons in dv QKD schemes are strongly
restricted by the today availably technical resources. For pursuing high
secret key rates, the continuous variable (cv) QKD has attracted extensive
interest in recent years \cite
{eight,nine,ten,eleven,twelve,thirteen,fourteen,fifteen,sixteen,seventeen,eighteen,nineteen,twenty}%
. In addition, another motivation to deal with cv QKD is that the mature
technology in quantum optics utilizing continuous quadrature amplitudes of
the quantized electromagnetic field, including the preparation, manipulation
and measurement of quantum states, can be used directly in cv quantum
communication to provide efficient implementation. Due to the lack of fast
and efficient single-photon detectors available at the moment, the nearly
unit quantum efficiency of bright-light photodetectors at high speeds
becomes the most attractive character for the explorers of cv QKD. On the
other hand, the unconditionalness of cv entangled states emerged from the
nonlinear optical interaction in an unconditional fashion is a valuable
feature. Earlier, in 1993 J. Kimble's group in Pasadena accomplished the
first experiment on quantum communication based on cv entanglement, in which
the idea about the quantum secure communication was involved although the
security of the protocol was not analyzed carefully \cite{twelve}. Later,
the cv entangled states has been successfully applied in various quantum
communication protocols to demonstrate the unconditional quantum
teleportation, entanglement swapping, quantum dense coding, and so on \cite
{twenty-one,twenty-two,twenty-three}.

A variety of cv QKD schemes employing the entanglement of the amplitude and
phase quadratures of optical fields were successively proposed. In the
scheme proposed by Ralph \cite{ten} the entangled EPR fields, which are
created by combining two independent amplitude-squeezed electromagnetic
fields, are sent to Bob along with their local oscillators. A random phase
shift is added to one of the fields to prevent an eavesdropper from
retrieving the information by simple interference at a beam-splitter. Reid
proposed a similar scheme exploiting the quadrature entangled fields, in
which the protection against eavesdropping is provided by observing a Bell
inequality violation \cite{eleven}. In Ref. [15], the quantum keys are
generated by measuring randomly the amplitude or the phase quadrature of one
of the entangled EPR optical beams. Any disturbance introduced by an
eavesdropper will degrade the correlations and hence possibly be detected.
K. Bencheikh et al. proposed a QKD protocol based on cv entanglement, in
which the quantum key generation is achieved by distributing each mode of
EPR entangled optical beams to sender and receiver who perform instantaneous
measurements of quadratures \cite{fourteen}. The quantum correlation between
the measurement outcomes are used to constitute the bits of the random
secret key. In these protocols, 50\% of the bits at least are rejected
because the base incompatibility and both of EPR entangled beams can be
intercepted. Obviously, for a given noisy channel the optimal individual
attack is to take a fraction of the transmitted signal beam, which equals
the line losses at sender's site and then send the remainder to receiver
through own lossless line. In this case, the eavesdropper is totally
undetected, and gets maximum possible information according to the
no-cloning theorem. Although in the above mentioned schemes the two-mode
entangling optical fields are utilized, both of the entangling modes can be
intercepted, so the security will not be more enhanced than that using a
single mode coherent state of light when the optimal attack is used, which
has been theoretically demonstrated in Ref. [16]. However, if only one of
the entangling modes can be attacked and the other one can be exploited by
the authorized receiver merely the entanglement will be helpful for
increasing the mutual information between Alice and Bob and nothing to Eve.
Based on this idea we propose a cv QKD scheme using the quadrature
entanglement of two modes in which the security is enhanced and the
transmission efficiency of the secret key is increased due to utilizing
quantum dense coding communication.

According to the condition of information theory for secure communication,
i.e., for enabling extraction of a secure key using the error correction
techniques and the privacy amplification \cite{twenty-four,twenty-five}, the
mutual information between Alice and Bob, $I_{AB}$, must exceed the
information that either of them shares with Eve (eavesdropper), $I_{AE}$ and
$I_{BE}$. It has been proved in Ref. [16] that the condition $I_{AB}>I_{AE}$
($I_{BE}$) is always violated for the transmission losses beyond 3 dB (50\%)
whatever the carriers of the information are the coherent, the squeezed or
the entangled states of light. In the successively completed cv QKD
experiments, the 3 dB loss limit was beaten by utilizing classical
techniques, such as the method of the reverse reconciliation \cite{eighteen}
or the post-selection procedure \cite{nineteen,twenty}. Is it possible to
beat the apparent 3 dB loss limit only using the presently available quantum
resources without demanding either the advanced and the unrealized quantum
techniques (quantum memories and entanglement purification et al.) or the
above mentioned classical methods? We note, in all communication schemes
discussed by Ref. [16] the information is transmitted unidirectionally from
Alice to Bob. In this case, the noise added in Alice's side cancels out
because it disturbs equally Eve and Bob. Therefore, the security of these
protocols does not rely on the noise feature of the used light beams.
Especially, in the nonmodulated cv QKD scheme using Einstein-Podolsky-Rosen
(EPR) entangled beams proposed by Silberhorn et al. \cite{fifteen}, Alice
keeps one of the EPR beams and sends the other to Bob. However, the
photocurrent signals detected by Alice are opened on the classical channel,
such that it is logically equivalent to a randomly modulated squeezed light
beam thus brings no any improvement of security \cite{sixteen}.

In this paper, we propose a round-trip transmission cv QKD scheme based on
EPR entanglement. A source producing EPR entangled light beams with the
quadrature amplitude and phase correlations is placed inside the station of
Bob. Only one of the EPR beams (signal beam) is sent to Alice and the other
one (idler beam) is retained by Bob and never opened even its classical
photocurrents. For enhancing transmission capacity we embed the benefits of
cv quantum dense coding \cite{twenty-six} in the cv QKD scheme as that done
by Degiovanni et al. at the dv protocol \cite{twenty-seven}. Two sets of
independent random numbers are modulated on the amplitude and phase
quadratures of Alice's EPR beam, respectively, then she sends the modulated
signal beam back to Bob. At Bob, the two sets of signals are simultaneously
decoded under the aid of the retained EPR beam with the homodyne detection
\cite{twenty-eight}. Since each half of the EPR beams has huge noise
individually and the correlation noises between a pair of EPR beams are
below the shot noise limit (SNL), if there is no too much excess noise added
in the beams the signal-to-noise ratios (SNR) in Alice's signal beam must be
lower than that decoded by Bob with the correlation measurements \cite
{twelve,twenty-three}. We found, in this scheme the condition of $%
I_{AB}>I_{AE}$ can be satisfied even when the transmission losses exceed 3
dB if higher EPR correlation is utilized. The security of the proposed
system is based on the determinative cv entanglement of EPR light beams and
the quantum no-cloning theorem. Two kinds of eavesdropper attack will be
discussed: 1. the individual quantum-tap attack to the modulated signal beam
on the way from Alice to Bob or to both unmodulated and modulated beams
using an optical beam-splitter according to the requiement of optimal
cloning. 2. the intercept-resend or partial intercept-mixing attack to the
non-modulated beam on the way from Bob to Alice using the simulated EPR
beams produced by Eve. The calculated results proved that the higher EPR
entanglement is helpful to beat the 3 dB loss limit in the proposed scheme.
The regions of the secure raw secret key, ($I_{AB}-I_{AE}$) $>0$, as the
functions of the line transmission and the EPR entanglement correlation
factor are calculated. Additionally, like the no-switching cv QKD scheme
presented in Refs. [17] and [20], the usual random switching between
measurement bases is not required and thus the channel capacity can be
significantly improved due to the application of quantum dense coding
method. Further, in the dense coding cv QKD scheme we utilize the bright EPR
beams with the anticorrelated amplitude quadratures and the correlated phase
quadratures as well as the Bell-state direct detection technique proposed
and used in our previous papers \cite{twenty-three,twenty-eight}, therefore
the local oscillation optical beam is not needed in the Bob's measurements
for extracting the secret keys. Naturally, the protocol dispenses with the
technical limitation on the communication bandwidth placed by the local
oscillator switching and is relatively simple to implement. Here we should
mention although the proposed protocol can beat the loss limit of 3 dB in
principle, however due to practical difficulties to produce quadrature
entangled light with high entanglement degree and highly susceptibility of
cv entanglement to loss the reachable communication distance has to be
strictly limited by the available entanglement quality. It has been
theoretically demonstrated that the entanglement of a cv resource, though
being degraded on a transmission line with loss never vanishes completely
for any degree of the loss, thus once the technologies of cv entanglement
distillation and purification are exploited the drawbacks will be possible
to be overcome \cite{twenty-nine,thirty}.

The paper is organized as follows. In the second section the cv QKD system
is described. The security against the quantum-tap attack and the
intercept-resend attack are discussed in the third and the fourth section,
respectively. A brief conclusion is given in the fifth section.

\section{ Cv QKD system using EPR entangled optical beams and round-trip
transmission}

The schematic of the proposed cv QKD system is shown in Fig. 1. The bright
EPR optical beams, $\hat{a}$ and $\hat{b}$, with the anticorrelated
amplitude quadratures and the correlated phase quadratures are produced from
an entanglement source. For experiments, this kind of EPR beams can be
obtained with a nondegenerate optical parametric amplifier (NOPA) below the
pump threshold operating in the state of deamplification, which has been
well described in Ref. [23]. The field annihilation operators, $\hat{a}$ and
$\hat{b}$, are expressed in terms of the amplitude ($\hat{X}_{a(b)}$) and
phase ($\hat{Y}_{a(b)}$) quadrature operators:

\[
\hat{a}=\hat{X}_a+i\hat{Y}_a,
\]

\begin{equation}
\hat{b}=\hat{X}_b+i\hat{Y}_b.
\end{equation}
The quadrature operators can be written as the sum of a steady state and a
fluctuating component

\[
\hat{X}=<\hat{X}>+\delta \hat{X},
\]

\begin{equation}
\hat{Y}=<\hat{Y}>+\delta \hat{Y},
\end{equation}
which have variances of $V(\hat{X})=\langle (\delta \hat{X})^2\rangle $ and $%
V(\hat{Y})=\langle (\delta \hat{Y})^2\rangle $. For the EPR entangled
optical beams produced from a NOPA operating at deamplification, the
variances for each individual beam ($\hat{a}$ or $\hat{b}$) and the
correlation variances are determined by the correlation factor $\gamma $ (or
called squeezing factor), which depends on the strength and the time of
parametric interaction \cite{eleven,sixteen},
\begin{equation}
V(\hat{X}_a)=V(\hat{Y}_a)=V(\hat{X}_b)=V(\hat{Y}_b)=(\gamma +1/\gamma )/2,
\end{equation}
\begin{equation}
V(\hat{X}_a+\hat{X}_b)=V(\hat{Y}_a-\hat{Y}_b)=2\gamma ,
\end{equation}
where we have assumed that the two modes, $\hat{a}$ and $\hat{b}$, are
totally balanced during the process of measurements and this requirement is
easily achieved in the experiments. The values of $\gamma $ are taken from 0
to 1, $\gamma =0$ and $\gamma =1$ correspond to the ideally perfect
correlation and no any correlation between $\hat{a}$ and $\hat{b}$,
respectively. The modes $\hat{a}$ and $\hat{b}$ from NOPA have the
orthogonal polarizations and can be separated with a
polarizing-beam-splitter (PBS). The entanglement source is placed inside the
station of Bob. The optical mode $\hat{a}$, say the signal mode, is sent to
Alice as the quantum channel of the transmitted signals. The mode $\hat{b}$,
say the idler mode, is retained by Bob and never is opened. The
beam-splitter R$_1$ and R$_2$ with same reflectivity (such as R = 10\%) are
placed in Bob and Alice who extract a small part from $\hat{b}$\ and $\hat{a}
$\ beam by means of R$_1$\ and R$_2$, respectively. The extracted beam from $%
\hat{b}$ ($\hat{a}$) is detected by the balanced-homodyne-detector HD$_1$ (HD%
$_2$) for checking the possible eavesdropper on the way from Bob to Alice,
which will be discussed in Section IV. Like the cv coherent state quantum
cryptography protocol presented in Refs. [18] and [20], for the optimal
information rate both amplitude and phase are modulated with Gaussian random
numbers \cite{twenty-nine}. Alice draws two random real numbers, $X_s$ and $%
Y_s$, from Gaussian distributions with zero mean and a variance of $V(X_s)$
and $V(Y_s)$. Then she modulates the amplitude and phase quadratures of the
transmitted $\hat{a}$ from R$_2$ by $X_s$ and $Y_s$ with the amplitude (AM)
and phase (PM) modulators, respectively. Alice transmits the modulated
signal mode $\hat{a}$ back to Bob. We assume that the channel transmission
efficiencies from Bob to Alice and from Alice to Bob without the presence of
Eve are identical and equal to $\eta $. Bob demodulates simultaneously the
modulated amplitude ($X_s$) and phase ($Y_s$) signals using the Bell-state
direct detection under the help of the retained mode $\hat{b}$ \cite
{twenty-eight}. The Bell-state direct detection system consists of a 50-50
beam-splitter (BS), a pair of photoelectric detectors (D$_1$\ and D$_2$),
two radio-frequency splitter (RF1 and RF2), a $\pi /2$\ phase shifter (PH),
a positive and a negative power combiner ($\oplus $\ and $\ominus $). The
beams $\hat{b}_1$\ and $\hat{a}_2$\ interfere on the beam-splitter (BS) and
then the output beams $\hat{c}$\ and $\hat{d}$\ are directly detected by D$%
_1 $\ and D$_2$, respectively. $\hat{\nu}$ stands for the vacuum noise added
in the quantum channel due to losses. The vacuum noises in different terms
are not correlated, thus they have to be considered independently. In the
followings we will calculate the physical conditions for satisfying $%
I_{AB}>I_{AE}$.

\section{Security conditions against optimal quantum-tap attack}

It has been theoretically demonstrated by Curty et al. \cite{thirty-one}
that the presence of detectable entanglement in a quantum state effectively
distributed between Alice and Bob is a necessary precondition for successful
key distillation. In the proposed scheme, the quantum entanglement of
quadratures shared by Alice and Bob is always existent and detectable so far
as the original EPR entanglement is not exhausted totally by the line
losses. The existence of cv entanglement between Alice and Bob provides the
base of security for the proposed protocol. The absolute theoretical
security of cv QKD protocols against any type of attack has already been
proven \cite{thirteen,thirty-two,thirty-three,thirty-four}. Here we do not
address the issue of unconditional security and also do not involve the
collective attack, which requires the quantum memory that is not easy to be
prepared in today technical condition. We consider security against
individual attacks only. Obviously, at the loss limit the optimal individual
attack is that Eve replaces the lossy channel by a perfect one with an
adapted beam splitter to mimic the losses and then generate a cloned signal
with a fidelity depending on the beam splitter transmission. In this case
Eve is totally not detected and gets the maximum possible information
according to the no-cloning theorem which is the maximum information amount
allowed by the laws of physics.{\bf \ }Since there is no any signal on the
quantum channel from Bob to Alice, we first consider a splitting attack on
the channel from Alice to Bob. A splitting attack involving both channels
will be considered later in this section.

We will use the Shannon formula of the optimum information rate to calculate
the raw secret key rate. The optimum mutual information ($I$) of a noisy
transmission channel is \cite{thirty-five}
\begin{equation}
I=(1/2)log_2(1+S/N),
\end{equation}
$S/N$ is the signal-to-noise ratio (SNR). Alice and Bob can establish a
secret key if and only if $I_{AB}>I_{AE}$, thus the secret key rate is
expressed as \cite{sixteen}
\begin{equation}
\Delta I=I_{AB}-I_{AE}.
\end{equation}
In the case of $\Delta I>0$ the communication between Alice and Bob will be
viewed as being secure. The simpler quantum-tap attack is to take a fraction
($1-\eta $) of the beam with the modulated signals at Alice's site, and to
send the fraction $\eta $ to Bob through her own lossless line. The
modulated optical mode at Alice is written as:
\begin{equation}
\hat{a}_1=\sqrt{1-R}(\sqrt{\eta }\hat{a}+\sqrt{1-\eta }\hat{\nu})+\sqrt{R}%
\hat{\nu}+s,
\end{equation}
and when the mode is transmitted back to Bob the mode $\hat{a}_1$ becomes $%
\hat{a}_2$ if without any interception:
\begin{eqnarray}
\hat{a}_2 &=&\sqrt{\eta }\hat{a}_1+\sqrt{1-\eta }\hat{\nu}  \nonumber \\
&=&\sqrt{1-R}[\eta \hat{a}+\sqrt{\eta (1-\eta )}\hat{\nu}]+\sqrt{R\eta }\hat{%
\nu}+\sqrt{1-\eta }\hat{\nu}+\sqrt{\eta }s,
\end{eqnarray}
where R is the reflectivity of the beam-splitter R$_1$ and R$_2$, $s$ stands
for the state of the modulated signals. The signal state is prepared at
Alice by displacing the amplitude and phase quadratures of a vacuum state by
$X_s$ and $Y_s$ respectively.

In order to measure the amplitude and phase signals by means of the
Bell-state direct detection simultaneously, Bob has to attenuate the
retained idler beam $\hat{b}$ to balance $\hat{a}_2$. The attenuated $\hat{b}
$ is expressed by $\hat{b}_1$:
\begin{equation}
\hat{b}_1=\sqrt{1-R}[\eta \hat{b}+\sqrt{\eta (1-\eta )}\hat{\nu}]+\sqrt{%
R\eta }\hat{\nu}+\sqrt{1-\eta }\hat{\nu}.
\end{equation}

The two output fields from the 50-50 beam-splitter (BS) of the Bell-state
direct detection system are \cite{twenty-eight}
\begin{equation}
\hat{c}=\frac 1{\sqrt{2}}(\hat{a}_2+i\hat{b}_1),
\end{equation}
\begin{equation}
\hat{d}=\frac 1{\sqrt{2}}(\hat{a}_2-i\hat{b}_1).
\end{equation}
The variances of the sum and the difference of $\hat{c}$ and $\hat{d}$ equal
to \cite{twenty-eight}:
\begin{equation}
V_{BX}=\frac 12[(1-R)\eta ^2V(\hat{X}_a+\hat{X}_b)+2(1-\eta ^2+R\eta
^2)+\eta V(X_s)],
\end{equation}
\begin{equation}
V_{BY}=\frac 12[(1-R)\eta ^2V(\hat{Y}_a-\hat{Y}_b)+2(1-\eta ^2+R\eta
^2)+\eta V(Y_s)],
\end{equation}
where, $V(\hat{X}_a+\hat{X}_b)$, $V(\hat{Y}_a-\hat{Y}_b)$ and $V(X_s)$, $%
V(Y_s)$ are the normalized correlation variances of the quadratures between $%
\hat{a}$ and $\hat{b}$ and the normalized variances of the signal
quadratures, respectively. The SNR of Bob's measurement for the amplitude
and the phase signals are respectively:
\begin{equation}
(S/N)_{BX}=\frac{\eta V(X_s)}{(1-R)\eta ^2V(\hat{X}_a+\hat{X}_b)+2(1-\eta
^2+R\eta ^2)},
\end{equation}
\begin{equation}
(S/N)_{BY}=\frac{\eta V(Y_s)}{(1-R)\eta ^2V(\hat{Y}_a-\hat{Y}_b)+2(1-\eta
^2+R\eta ^2)}.
\end{equation}
For the dense coding QKD scheme, Eve takes a fraction ($1-\eta $) of the
beam $\hat{a}_1$\ at Alice's site and then simultaneously measures the
amplitude and phase quadratures of the intercepted beam by means of a 50-50
beam-splitter and two sets of homodyne detector (HD) as shown in Fig. 2. The
variances of the amplitude and phase quadratures measured by Eve are
respectively expressed by:
\begin{equation}
V_{EX}=\frac 12[\eta (1-\eta )(1-R)V(\hat{X}_a)+2-(1-R)\eta +(1-R)\eta
^2+(1-\eta )V(X_s)],
\end{equation}
\begin{equation}
V_{EY}=\frac 12[\eta (1-\eta )(1-R)V(\hat{Y}_a)+2-(1-R)\eta +(1-R)\eta
^2+(1-\eta )V(Y_s)].
\end{equation}
The SNR of Eve's measurements equal to
\begin{equation}
(S/N)_{EX}=\frac{(1-\eta )V(X_s)}{(1-R)\eta (1-\eta )V(\hat{X}%
_a)+2-(1-R)\eta +(1-R)\eta ^2},
\end{equation}
\begin{equation}
(S/N)_{EY}=\frac{(1-\eta )V(Y_s)}{(1-R)\eta (1-\eta )V(\hat{Y}%
_a)+2-(1-R)\eta +(1-R)\eta ^2}.
\end{equation}
Substitute the equations (14) (15) and (18) (19) into the Eq. (5) we can
calculate the mutual information $I_{AB}^X$, $I_{AB}^Y$ and $I_{AE}^X$, $%
I_{AE}^Y$, as well as the total secret key rate $\Delta I$, which is the sum
of the secret key rates of the amplitude ($\Delta I^X$) and phase ($\Delta
I^Y$) quadrature
\begin{eqnarray}
\Delta I &=&I_{AB}^X+I_{AB}^Y-I_{AE}^X-I_{AE}^Y  \nonumber \\
&=&I_{AB}^X-I_{AE}^X+I_{AB}^Y-I_{AE}^Y  \nonumber \\
&=&\Delta I^X+\Delta I^Y.
\end{eqnarray}

Figure 3 shows the secret key rate as a function of the channel efficiency $%
\eta $ and the correlation factor $\gamma $ for the given signal variances
and the reflectivity R, where the normalized $V(X_s)=V(Y_s)=10$ and $R=0.1$.
In the case of $\gamma =1$, i.e. without any EPR correlation between $\hat{a}
$ and $\hat{b}$, the scheme is secure ($\Delta I>0$) only at $\eta >0.5$,
which corresponds to the conclusion in Ref. [16] for the unidirect
transmission. For comparison, the function of $\Delta I$ versus $\eta $ in
the unidirectional transmission (see Eq. (5) of Ref. [16]) is drawn in Fig.
3 with the dashed line. We can see, when $\eta >0.5$ our scheme is better
than that of the unidirectional transmission even $\gamma =1$ because the
dense coding scheme is applied. However, for the lower $\eta $ ($\eta <0.5$%
), the unidirectional transmission is advantaged since the
transmission losses are doubled in the round-trip transmission
protocol. For $\gamma =0.05 $ and $\gamma =0.4$ (corresponding to
the correlation degree of 13 dB and 4 dB, respectively), $\Delta I$
will be larger than zero once $\eta >0.33$ and $\eta >0.46$,
respectively. It means, the limitation of 3 dB losses ($\eta >0.5$)
can be beaten in our scheme using EPR entanglement if Eve only taps
the modulated channel. It is obvious from the equations of SNR, for
the channel with very high losses ($\eta \rightarrow 0$) the SNR
will only depend on the variances of the modulated
signal, $V(X_s)$ or $V(Y_s)$, thus all figures cross to the same point on $%
\Delta I$ axis [$\Delta I=-2.58$ in the case of $V(X_s)=V(Y_s)=10$].

Figure 4 shows the dependences of the secret key rate on the correlation
factor $\gamma $ for $\eta =$ 0.7, 0.5 and 0.25 respectively. The three
dashed lines correspond to the function curves of the unidirectional
transmission with $\eta =$ 0.7, 0.5 and 0.25 also for comparison. When $\eta
>0.5$, the secret key rates of the proposed scheme are always larger than
that of the unidirectional transmission. For $\eta =0.25$, the
unidirectional transmission is not secure always while our scheme can be
secure if the EPR correlation is high enough ($\gamma <0.02$).

An other better clone attack scheme is that Eve taps both the unmodulated
optical beam before Alice and the modulated signal beam after Alice, then to
substract the two intercepted signal beams for performing the correlation
measurement. In this case the two extracted beams are the thermal optical
fields and very noisy due to the added vacuum noise on the beam-splitter.
According to the requirement of the above-mentioned optimally individual
attack, the extracted amount should match the line loss. Eve has to
simultaneously measure the amplitude and phase quadratures of both the
intercepted unmodulated optical beam and the modulated signal beam by means
of four sets of homodyne detection systems. The measured amplitude and phase
quadratures of the intercepted light beams on the ways before and after
Alice are respectively:
\begin{equation}
\hat{X}_{E_1}^{^{\prime }}=\frac 1{\sqrt{2}}(\sqrt{1-\eta }\hat{X}_a-\sqrt{%
\eta }\hat{X}_\nu +\hat{X}_\nu ),
\end{equation}
\begin{equation}
\hat{Y}_{E_1}^{^{\prime }}=\frac 1{\sqrt{2}}(\sqrt{1-\eta }\hat{Y}_a-\sqrt{%
\eta }\hat{Y}_\nu -\hat{Y}_\nu );
\end{equation}
and
\begin{equation}
\hat{X}_{E_2}^{^{\prime }}=\frac 1{\sqrt{2}}[\sqrt{\eta (1-\eta )(1-R)}\hat{X%
}_a+(1-\eta )\sqrt{1-R}\hat{X}_\nu +\sqrt{R(1-\eta )}\hat{X}_\nu -\sqrt{\eta
}\hat{X}_\nu +\hat{X}_\nu +\sqrt{1-\eta }X_s],
\end{equation}
\begin{equation}
\hat{Y}_{E_2}^{^{\prime }}=\frac 1{\sqrt{2}}[\sqrt{\eta (1-\eta )(1-R)}\hat{Y%
}_a+(1-\eta )\sqrt{1-R}\hat{Y}_\nu +\sqrt{R(1-\eta )}\hat{Y}_\nu -\sqrt{\eta
}\hat{Y}_\nu -\hat{Y}_\nu +\sqrt{1-\eta }Y_s].
\end{equation}
For eliminating the thermal-like component of the noises [$V(\hat{X}_a)$\
and $V(\hat{Y}_a)$], Eve multiplies the tapped unmodulated signal beam [Eqs.
(21) and (22)] by $\sqrt{\eta (1-R)}$, that is to attenuate the optical beam
by a factor of $\sqrt{\eta (1-R)}$, then subtracts the tapped modulated
signal beam [Eqs. (23) and (24)] from it. The variances of [$\sqrt{\eta (1-R)%
}\hat{X}_{E_1}^{^{\prime }}-\hat{X}_{E_2}^{^{\prime }}$] and [$\sqrt{\eta
(1-R)}\hat{Y}_{E_1}^{^{\prime }}-\hat{Y}_{E_2}^{^{\prime }}$] equal to

\begin{equation}
V_{EX}^{^{\prime }}=\frac 12\{[\eta \sqrt{1-R}-(1-\eta )\sqrt{1-R}]^2+1+\eta
(1-R)+\eta +R(1-\eta )+(1-\eta )V(X_s)\},
\end{equation}

\begin{equation}
V_{EY}^{^{\prime }}=\frac 12\{[\eta \sqrt{1-R}-(1-\eta )\sqrt{1-R}]^2+1+\eta
(1-R)+\eta +R(1-\eta )+(1-\eta )V(Y_s)\}.
\end{equation}
The corresponding signal to noise ratios are

\begin{equation}
(S/N)_{EX}^{^{\prime }}=\frac{(1-\eta )V(X_s)}{[\eta \sqrt{1-R}-(1-\eta )%
\sqrt{1-R}]^2+1+\eta (1-R)+\eta +R(1-\eta )},
\end{equation}

\begin{equation}
(S/N)_{EY}^{^{\prime }}=\frac{(1-\eta )V(Y_s)}{[\eta \sqrt{1-R}-(1-\eta )%
\sqrt{1-R}]^2+1+\eta (1-R)+\eta +R(1-\eta )}.
\end{equation}
Figure 5 shows the function curves of the secret key rate $\Delta I$\ versus
the channel efficiency $\eta $\ for the dual-tap attack, where the
normalized $V(X_s)=V(Y_s)=10$\ and $R=0.1$. For $\gamma =1$, the secret key
rate is larger than zero only when $\eta >0.5$, which is same to the case of
only tapping the modulated signal. For $\gamma =0.05$\ and $\gamma =0.4$,
the secret key rate is positive when $\eta >0.46$\ and $\eta >0.474$,
respectively. We see, under this attack due to that the influence of the
thermal-like component of noises involved in the two intercepted optical
beams is canceled, the effect of increasing EPR correlation to the secret
key rate will not be as strong as that only attacking modulated signal beam
(Fig. 3 and 4). However, we still see the possibility to beat the apparent 3
dB loss limit of the raw secret key rate.

\section{Security against intercept-resend attack}

An other possible attack on the way from Bob to Alice is that Eve intercepts
totally the signal beam without the modulated signals and retains it in her
station, then she sends a simulated beam, which can be a half of simulated
EPR entangled beams, to Alice. If Alice does not know that the beam is
simulated, she will modulate the signals on it and send it out as usual. Eve
intercepts the modulated signal beam again and demodulates the signals using
the other half of the simulated EPR beams retained by herself as that done
by Bob. At last she modulates the same signals on the real signal beam and
sends it back to Bob. Fortunately, in our scheme Bob retains a half of the
real EPR beams and never opens it. Any simulated beam can not be
quantum-correlated with the real one. Bob and Alice take a part of the beam $%
\hat{b}$ and $\hat{a}$ from R$_1$ and R$_2$ respectively, then randomly
measure the amplitude or phase quadrature of the taken partial beam with the
homodyne detector (HD$_1$ and HD$_2$) during the communication. Alice sends
her measurement results to Bob by a classical channel and Bob checks the
quantum correlation between his quadratures and Alice's quadratures measured
simultaneously. The correlation variances of the amplitude ($V_{RX}$) and
the phase ($V_{RY}$) between the reflected beams from R$_1$ and R$_2$ are:

\begin{equation}
V_{RX}=\frac 12R\eta V(\hat{X}_a+\hat{X}_b)+1-R\eta ,
\end{equation}

\begin{equation}
V_{RY}=\frac 12R\eta V(\hat{Y}_a-\hat{Y}_b)+1-R\eta .
\end{equation}
Statistically, they simultaneously measure the same quadrature with
50\% probability and the correlation variances of the two measured
values will be below the shot noise limit (SNL) due to the existence
of EPR correlations between $\hat{a}$ and $\hat{b}$ if there is no
the intercept -resend attack. When the simulated beam is used, the
quantum correlation disappear thus the variances always are higher
than the SNL. The quantum noncloning forbids Eve to copy the quantum
fluctuation of the real signal beam and provides the physical
mechanism of security for our scheme. In cv experiments the
correlation variances of quadratures up to 0.1 dB below the SNL can
be precisely measured by means of homodyne detectors under today
technical condition. As a loose limitation we take 0.3 dB, which was
detected in the early experiment of quantum optics in 1985 \cite{thirty-six}%
, to be a bound for detecting Eve with a high probability over 99\%. For
given original EPR entanglement and the line transmission, the requirement $%
V_{RX}=V_{RY}=$0.928 (according to 0.32 dB below the SNL) is reached when

\begin{equation}
R=\frac{2(V_{RX}-1)}{\eta [V(\hat{X}_a+\hat{X}_b)-2]}
\end{equation}
For example, if $\gamma =$\ 0.2 [corresponding to $V(\hat{X}_a+\hat{X}_b)=V(%
\hat{Y}_a-\hat{Y}_b)=$0.4, which is 7 dB below the SNL, which is
experimentally reachable at present \cite{thirty-seven}.], $\eta =$0.9\ and $%
V_{RX}=V_{RY}=$0.928, we have R $=$\ 0.1. In the calculations of equations
(7)-(20) the effect of R has been involved. The procedure checking the
intercept-resend attack can be performed instantaneously during the
communication proceeding and the communication will not be disturbed only
the original quantum correlation is decreased, for example if R = 0.1 the
correlation is decreased from 7 dB to 5.5 dB.

If Eve partially mixes her own EPR beam with the transferred real one
instead of totally replacing it, according to the optimally cloning scheme
the replaced partial amplitude should equal $\sqrt{1-\eta }\hat{X}_e(\hat{Y}%
_e)$ [$\hat{X}_e(\hat{Y}_e)$\ is the amplitude (phase) of a half of EPR
beams prepared by Eve and $\left| \hat{X}_e\right| =\left| \hat{X}_a\right|
=\left| \hat{Y}_e\right| =\left| \hat{Y}_a\right| $]. The amplitude and
phase quadratures of the optical field received by Alice are

\begin{equation}
\hat{X}_A^{^{\prime \prime }}=\sqrt{\eta }\hat{X}_a+\sqrt{1-\eta }\hat{X}_e,
\end{equation}

\begin{equation}
\hat{Y}_A^{^{\prime \prime }}=\sqrt{\eta }\hat{Y}_a+\sqrt{1-\eta }\hat{Y}_e,
\end{equation}
and the remainders retained by Eve are

\begin{equation}
\hat{X}_E^{^{\prime \prime }}=\sqrt{1-\eta }\hat{X}_a-\sqrt{\eta }\hat{X}_e,
\end{equation}

\begin{equation}
\hat{Y}_E^{^{\prime \prime }}=\sqrt{1-\eta }\hat{Y}_a-\sqrt{\eta }\hat{Y}_e.
\end{equation}
Then Eve intercepts all modulated beam and performs the correlation
measurement using the other half ($\hat{X}_f,\hat{Y}_f$)of the entangled
beams prepared by Eve. In this case, $V(\hat{X}_e+\hat{X}_f)$[$V(\hat{Y}_e-%
\hat{Y}_f)$]\ should be smaller than SNL and depends on the correlation
degree ($\gamma _E$) of Eve's EPR beams. The correlation variances measured
by Eve equal to

\begin{equation}
V_{EX}^{^{\prime \prime }}=\frac 12[(1-R)(1-\eta )V(\hat{X}_e+\hat{X}%
_f)+(1-R)\eta V(\hat{X}_a)+(1-R)\eta +2R+V(X_s)],
\end{equation}

\begin{equation}
V_{EY}^{^{\prime \prime }}=\frac 12[(1-R)(1-\eta )V(\hat{Y}_e-\hat{Y}%
_f)+(1-R)\eta V(\hat{Y}_a)+(1-R)\eta +2R+V(Y_s)].
\end{equation}
The corresponding SNRs are

\begin{equation}
(S/N)_{EX}^{^{\prime \prime }}=\frac{V(X_s)}{(1-R)(1-\eta )V(\hat{X}_e+\hat{X%
}_f)+(1-R)\eta V(\hat{X}_a)+(1-R)\eta +2R},
\end{equation}

\begin{equation}
(S/N)_{EY}^{^{\prime \prime }}=\frac{V(Y_s)}{(1-R)(1-\eta )V(\hat{Y}_e-\hat{Y%
}_f)+(1-R)\eta V(\hat{Y}_a)+(1-R)\eta +2R}.
\end{equation}

The correlation degree between Bob's beam reflected from R$_1$\ and Alice's
beam from R$_2$\ must be decreased since the partial non-correlated light [$%
\hat{X}_e(\hat{Y}_e)$] is mixed in the measured beams. The calculated
variances are

\begin{equation}
V_{RX}^{^{\prime \prime }}=\frac 12[R\eta V(\hat{X}_a+\hat{X}_b)+R(1-\eta )V(%
\hat{X}_e)+2-R-R\eta ],
\end{equation}

\begin{equation}
V_{RY}^{^{\prime \prime }}=\frac 12[R\eta V(\hat{Y}_a-\hat{Y}_b)+R(1-\eta )V(%
\hat{Y}_e)+2-R-R\eta ].
\end{equation}
The higher the EPR correlation between $\hat{X}_e$($\hat{Y}_e$)\ and $\hat{X}%
_f$($\hat{Y}_f$) is, the larger the $V(\hat{X}_e)$[$V(\hat{Y}_e)$]\ is, thus
$V_{RX}^{^{\prime \prime }}$($V_{RY}^{^{\prime \prime }}$)\ increases. For
example, taking $\gamma _E=$0.05\ (corresponding to 13 dB below the SNL),
when $\gamma =$0.2\ (7 dB below the SNL), $V(X_s)=V(Y_s)=$10, $R=$0.1\ and $%
\eta =$0.9, if the light intensity mixed by Eve is 10\% of the total
intensity, the correlation degree between Alice's and Bob's reflected beams
will be reduced from 0.32 dB (without eavesdropper) to 0.11 dB, thus the
presence of Eve will be revealed.

On the other hand, Eve has to add partial uncorrelation light into the beam
retained by her to make Bob receiving equal intensity of light, otherwise
she must be revealed immediately. The amplitude and phase quadratures of the
light beam sent back to Bob are

\begin{equation}
\hat{X}_a^{^{\prime \prime }}=\eta \sqrt{1-R}\hat{X}_a-\frac{\eta \sqrt{1-R}%
}{\sqrt{1-\eta }}\sqrt{\eta }\hat{X}_e+\sqrt{1-\frac{\eta ^2(1-R)}{1-\eta }}%
\hat{X}_\nu +\sqrt{\eta }X_s,
\end{equation}

\begin{equation}
\hat{Y}_a^{^{\prime \prime }}=\eta \sqrt{1-R}\hat{Y}_a-\frac{\eta \sqrt{1-R}%
}{\sqrt{1-\eta }}\sqrt{\eta }\hat{Y}_e+\sqrt{1-\frac{\eta ^2(1-R)}{1-\eta }}%
\hat{Y}_\nu +\sqrt{\eta }Y_s.
\end{equation}
The correlation variances and the SNRs measured by Bob equal to

\begin{equation}
V_{BX}^{^{\prime \prime }}=\frac 12[(1-R)\eta ^2V(\hat{X}_a+\hat{X}_b)+\frac{%
\eta ^3(1-R)}{1-\eta }V(\hat{X}_e)+2-(1-R)\eta ^2-\frac{\eta ^2(1-R)}{1-\eta
}+\eta V(X_s)],
\end{equation}

\begin{equation}
V_{BY}^{^{\prime \prime }}=\frac 12[(1-R)\eta ^2V(\hat{Y}_a-\hat{Y}_b)+\frac{%
\eta ^3(1-R)}{1-\eta }V(\hat{Y}_e)+2-(1-R)\eta ^2-\frac{\eta ^2(1-R)}{1-\eta
}+\eta V(Y_s)],
\end{equation}
and
\begin{equation}
(S/N)_{BX}^{^{\prime \prime }}=\frac{\eta V(X_s)}{(1-R)\eta ^2V(\hat{X}_a+%
\hat{X}_b)+\frac{\eta ^3(1-R)}{1-\eta }V(\hat{X}_e)+2-(1-R)\eta ^2-\frac{%
\eta ^2(1-R)}{1-\eta }},
\end{equation}
\begin{equation}
(S/N)_{BY}^{^{\prime \prime }}=\frac{\eta V(Y_s)}{(1-R)\eta ^2V(\hat{Y}_a-%
\hat{Y}_b)+\frac{\eta ^3(1-R)}{1-\eta }V(\hat{Y}_e)+2-(1-R)\eta ^2-\frac{%
\eta ^2(1-R)}{1-\eta }}.
\end{equation}
If Eve replaces 10\% of total beam intensity the SNRs calculated with Eq.
(46) and (47) are $(S/N)_{BX}^{^{\prime \prime }}=(S/N)_{BY}^{^{\prime
\prime }}=$0.15 ( $\gamma _E=$0.05, $\gamma =$0.2, $V(X_s)=V(Y_s)=$10, $R=$%
0.1\ and $\eta =$0.9) and the SNRs calculated with Eq. (14) and (15), which
only consider the optimal cloning attack of one channel, are $%
(S/N)_{BX}=(S/N)_{BY}=$10.8 with same parameters. The significant reduction
of the SNRs will clearly reveal the presence of Eve also.

Although Eve might obtained more information by means of the correlation
measurement using the EPR beams prepared by herself, her presence will be
also revealed. Of course, if Eve has perfect entangled beams and perfect
quantum memory the eavesdropping scheme might be better than that totally
intercepting. For more detailed discussion we have to compare the
information amounts respectively obtained by Bob and Eve in this case and
find the region for the secure transmission, which have been over the range
of this paper.

\section{Conclusion}

We proposed a round-trip transmission cv QKD scheme based on the EPR
entanglement of optical beams. We mixed the advantage of cv dense coding
into QKD, thus the secret key rate is significantly improved. Due to Bob's
simultaneously measuring both amplitude and phase quadratures the randomly
switching between measurement bases is not required, such that the serious
technical limitation on the communication bandwidth placed by the local
oscillator switching and the technical difficulty of precisely controlling
the phase of a local oscillator are not existent any more.

At last, we should mention, although the EPR optical beams with the
anticorrelated amplitude and correlated phase quadratures are discussed in
the paper, the scheme and all calculations are appropriate also to the EPR
optical beams with the correlated amplitude and anticorrelated phase
quadratures ($V(\hat{X}_a-\hat{X}_b)<$ SNL, $V(\hat{Y}_a+\hat{Y}_b)<$ SNL),
which can be produced from an optical parametric amplifier operating at
amplification \cite{twelve,twenty-one}. In this case, Bob only needs to
substitute the Bell-state direct detection with two sets of the normal
balanced homodyne detector as described in the original cv dense coding
paper \cite{twenty-six}.

Several possible attack schemes have been discussed, however the
unconditional security for the proposed protocol has not been demonstrated.
Although Eve's eavesdropping technologies are various, the discussed attack
schemes are representative and usually used. Especially, the discussions
about beam-splitter attack are based on the optimal individual attack
allowed by the no-cloning theorem of quantum mechanics, thus have common
significance. The theoretical demonstration of unconditional security for
the proposed scheme is beyond the scope of the paper and might be remained
to be an open question.

Due to the sensitivity of the optical quantum entanglement to losses the
application of the proposed scheme in the long-distance communication is
limited. However the proposed scheme have shown the possiblility to beat the
loss limit of 3 dB using the cv EPR entanglement. Along with the development
of quantum optical technology the higher and higher determinative cv
entanglement can be reached, so the potential of the proposed scheme in
future application is expected.

This work was supported by the National Natural Science Foundation of China
(Grants No. 60238010 and 60378014).

*Email address: panqing@sxu.edu.cn

\smallskip

Captions of figures:

Fig. 1 The schematic of the quantum key distribution with bright EPR beams.

AM-amplitude modulator, PM-phase modulator, R-reflection rate of the
beam-splitter R$_1$ and R$_2$, PBS-polarization beamsplitter, HD-homodyne
detection system, $\eta $-channel efficiency, PH-$\pi /2$\ phase shifter,
RF-radio-frequency splitter.

Fig. 2 Eve's detection system. HD-homodyne detection system, $\eta $-channel
efficiency.

Fig. 3 The secret key rate versus channel efficiency at various correlation
degrees for the case only tapping modulated beam. $V(X_s)=V(Y_s)=10,$ $R=0.1$%
, solid lines-the function curves of the presented scheme, dashed line-the
function curves of the unidirectional transmission scheme.

Fig. 4 The secret key rate versus correlation degree at various channel
efficiencies for the case only tapping modulated beam. $V(X_s)=V(Y_s)=10,$ $%
R=0.1$, solid lines-the function curves of the presented scheme, dashed
lines-the function curves of the unidirectional transmission scheme.

Fig. 5 The secret key rate versus channel efficiency at various correlation
degrees for the case tapping both modulated and unmodulated beams. $%
V(X_s)=V(Y_s)=10,$ $R=0.1$

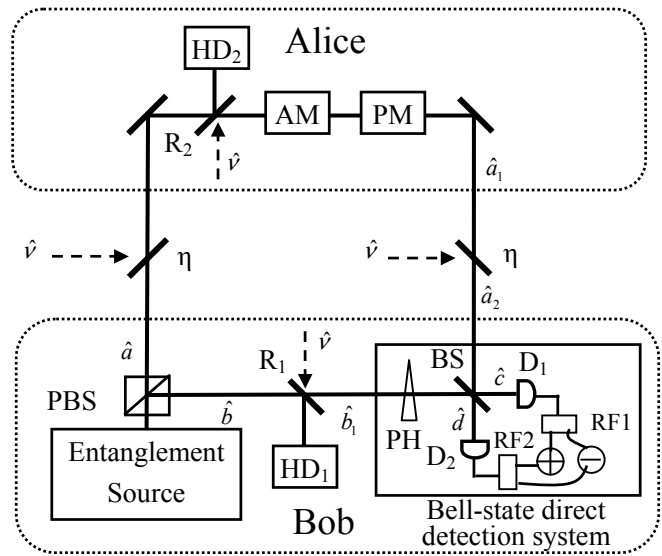
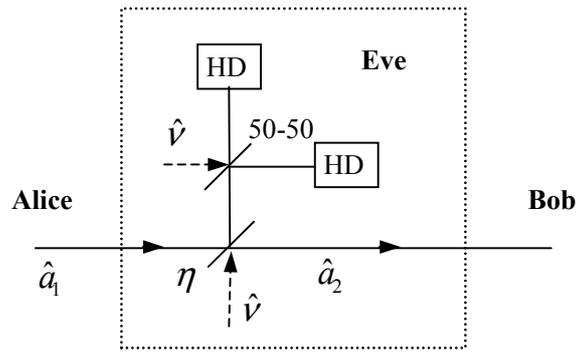
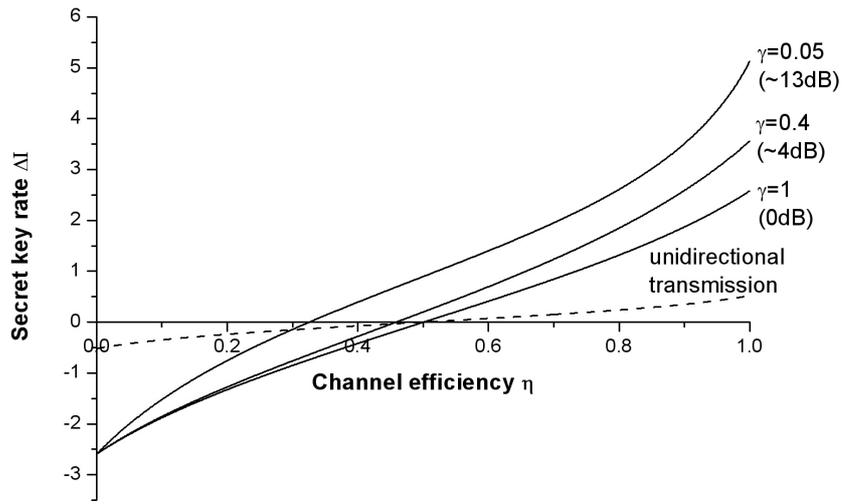
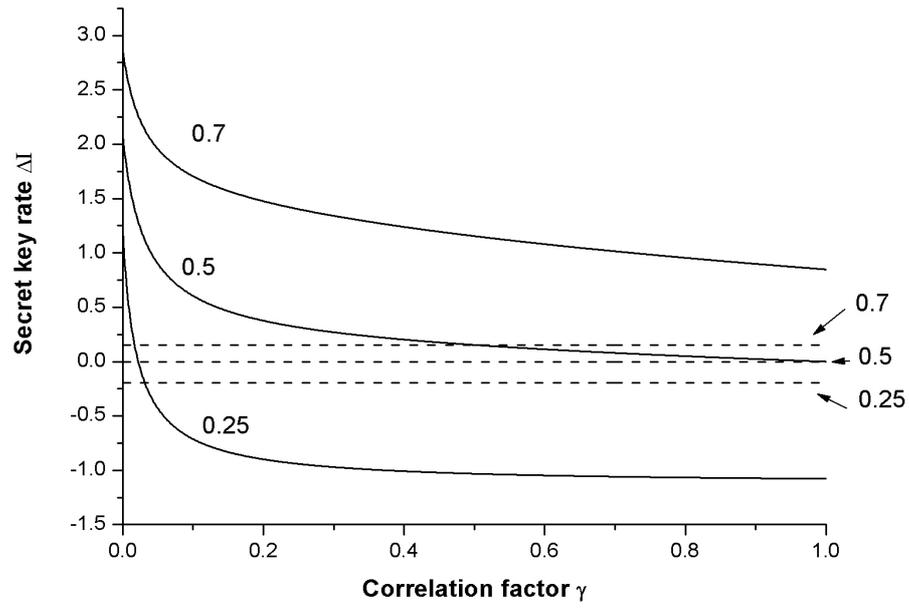
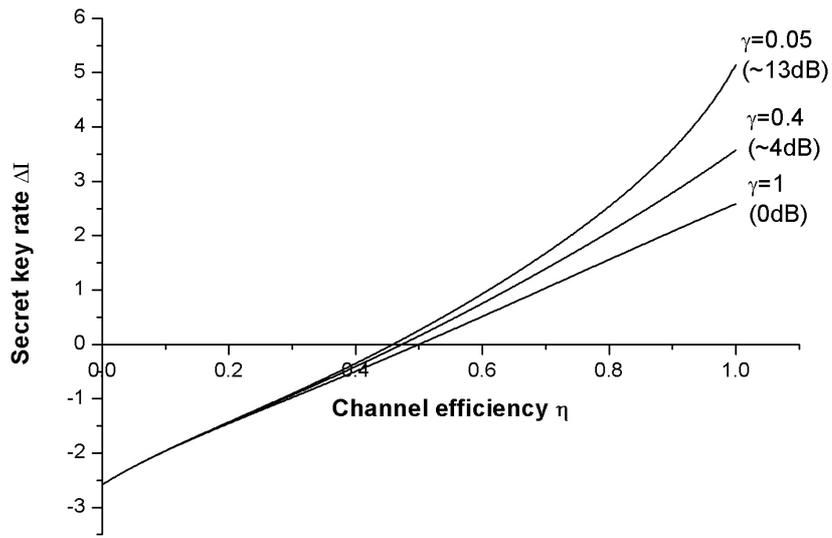
\end{document}